\documentclass[aps,prl,twocolumn,groupedaddress,amssymb,amsfonts,showpacs]{revtex4}
\usepackage{times}

\usepackage{color}
\usepackage[dvips]{graphics,graphicx}

\usepackage{calc,ifthen}

\newlength{\elimdepthdim}
\newlength{\elimheightdim}
\newlength{\elimwidthdim}

\newlength{\strutdepthdim}
\newlength{\strutheightdim}
\newlength{\strutwidthdim}

\def\one{{\mathchoice {\rm 1\mskip-4mu l} {\rm 1\mskip-4mu l} {\rm
1\mskip-4.5mu l} {\rm 1\mskip-5mu l}}}

\newcommand{\ket}[1]{\qvbar{#1}\qrangle}
\newcommand{\bra}[1]{\qlangle{#1}\qvbar}

\newcounter{herefignum}

\newcommand{\shortqph}[1]{}

\providecommand{\ignore}[1]{}

\newcommand{\mCo}[1]{\textcolor{blue}{}}

\newcommand{\hC}[1]{\textcolor{red}{}}

\def\openone{\leavevmode\hbox{\small1\kern-3.8pt\normalsize1}}
\def\RR{{\rm I\kern-.2emR}}
\def\tr{{\rm tr}\; }

\def\fsu{\mathfrak{su}}

\def\openone{\leavevmode\hbox{\small1\kern-3.8pt\normalsize1}}
\def\RR{{\rm I\kern-.2emR}}
\def\tr{{\rm tr}\; }

\def\hA{\hat{A}}

\providecommand{\ignore}[1]{}

\renewcommand{\ket}[1]{| #1 \rangle}
\renewcommand{\bra}[1]{\langle #1 |}

\newcommand{\bitem}{\begin{itemize}}
\newcommand{\eitem}{\end{itemize}}
\newcommand{\benum}{\begin{enumerate}}
\newcommand{\eenum}{\end{enumerate}}
\newcommand{\beq}{\begin{equation}}
\newcommand{\eeq}{\end{equation}}
\newcommand{\beqa}{\begin{eqnarray}}
\newcommand{\eeqa}{\end{eqnarray}}
\newtheorem{definition}{Definition}

\newtheorem{proposition}{Proposition}

\newcommand{\bproof}{\begin{proof}}
\newcommand{\eproof}{\end{proof}}
\newcommand{\bprop}{\begin{proposition}}

\newcommand{\bdef}{\begin{definition}}

\begin{document}

%\preprint{}

\title{A Quantum Approach to Classical Statistical Mechanics}


%\author{Howard Barnum$^1$, Emanuel Knill$^1$, Gerardo Ortiz$^2$,
%Rolando Somma$^1$}
%\affiliation{$^1$CCS-3, Mail Stop B256, Los Alamos National Laboratory,
%Los Alamos, NM 87545\\ $^2$T-11, Mail Stop B262, Los Alamos National 
%Laboratory, Los Alamos, NM 87545}
%\thanks{}
%\affiliation{}

\author{R. D. Somma}
\email{somma@lanl.gov}

\affiliation{Los Alamos National Laboratory, Los Alamos, NM 87545, USA}
\author{C. D. Batista}
\affiliation{Los Alamos National Laboratory, Los Alamos, NM 87545, USA}
%\email{somma@lanl.gov,barnum@lanl.gov,g_ortiz@lanl.gov,knill@boulder.nist.gov}
%\email{barnum@lanl.gov}
\author{G. Ortiz}
\affiliation{Department of Physics, Indiana University, Bloomington, IN
47405, USA}
%\email{somma@lanl.gov,barnum@lanl.gov,g_ortiz@lanl.gov,knill@boulder.nist.gov}

%\author{Emanuel Knill}
%\email{somma@lanl.gov,barnum@lanl.gov,g_ortiz@lanl.gov,knill@boulder.nist.gov}

%\affiliation{$^1$Los Alamos National Laboratory, MS D454, Los Alamos, NM 87545}

%\affiliation{Mathematical and Computational Sciences Division, National
%Institute of Standards and Technology, Boulder CO 80305}
%\email{knill@boulder.nist.gov}
%\email{g_ortiz@lanl.gov}


\date{\today}
\begin{abstract}
We present a new approach to study the thermodynamic properties of
$d$-dimensional classical systems by reducing the problem to the
computation of ground state properties of a $d$-dimensional quantum
model. This classical-to-quantum mapping allows us to deal with
standard  optimization methods, such as simulated and quantum
annealing, on  an equal basis. Consequently, we extend the quantum
annealing method to simulate classical systems at finite
temperatures.  Using the adiabatic theorem of quantum mechanics,  we
derive the rates to assure convergence to the optimal
thermodynamic state. For simulated and quantum annealing, we obtain the
asymptotic  rates of $T(t) \approx (p N) /(k_B \log t)$ and $\gamma(t) \approx 
(Nt)^{-\bar{c}/N}$, for the temperature and magnetic field, respectively.
Other annealing strategies, as well as their potential speed-up, are
also discussed.
\end{abstract}


\maketitle

An outstanding issue in combinatorial optimization is the
classification of problems according to their computational complexity.
Typically, one defines a cost function that needs to be minimized and
the question is how the number of resources (e.g., time) to determine
the minimum  scales with the problem size $N$. Long time ago it has
been recognized that certain physics problems can be cast in this
language. For example,  it has been shown that the computation of the
ground state energy (or the partition function) of classical
three-dimensional spin glasses belongs to the class of NP-complete
problems~\cite{bar82}, i.e. there is no known algorithm that can find
the solution with polynomial (in $N$) resources. After all, the number
of possible microscopic configurations  of the system increases
exponentially with the system size $N$ and, unless certain {\it
symmetries} reduce the complexity, one has to search in an
exponentially  large state space. This simplification happens, for
example, in the two-dimensional Ising spin glass~\cite{ons44} (or any
planar graph or lattice). 

Simulated annealing (SA)~\cite{kir83} and quantum annealing 
(QA)~\cite{das05} represent general algorithmic strategies to attack
these optimization problems. The basic idea consists in finding the
solution to the optimization problem as a limit of an effective
physical process which uses additional variables or dimensions, and
where the cost function is  identified with a Hamiltonian $H$ of a {\it
classical} physical system. In SA one  introduces temperature $T$ as a
tunable parameter: Initially the system is heated and next $T(t)$ is
slowly decreased towards zero, eventually converging to the ground
(lowest energy) state, whose energy equals the cost function. In QA,
however, a time-dependent {\it ad-hoc} external magnetic field of
magnitude $\gamma(t)$ is added to $H$, such that the total Hamiltonian
can be interpreted as that of a quantum system.  The (quantum)
annealing process consists of slowly decreasing $\gamma(t)$  from a
large value towards zero, while keeping $T=0$. Since a quantum system
in $d$ dimensions can be mapped onto another classical system in
$d+1$~\cite{pol93},  effectively in QA one is adding one extra {\it
space}-dimension to the problem.  In both strategies, the annealing
procedure is essential to converge to the desired (ground) state, as it
avoids getting stuck in local minima,  using less resources than other
optimization methods~\cite{har02}.
 
In this Letter, we propose new algorithms to study the thermodynamic
properties of classical systems (including frustrated systems, such as
spin glasses). The crux of the method consists of mapping the classical
$d$-dimensional problem into a quantum problem of the same
dimensionality, and then using techniques similar  to those of QA to
solve the latter. Our particular mapping allows us to unify  the
methods of SA and QA, and extend them to:  i) study arbitrary classical
models at $T >0$ and ii) study new annealing schemes. From this
classical-quantum mapping perspective, any annealing strategy differs
by the choice of path in (quantum) Hamiltonian space. Computation of
thermodynamic properties of the classical model amounts then to
computation of ground state properties of the mapped quantum model. Our
approach can be readily implemented on a classical computer (CC) by
using existent stochastic methods, such as Green's Function Monte
Carlo~\cite{kal74}, or by simulating the corresponding time-dependent
Schr\"odinger equation. Since the proposed algorithms are based on a
slow change of  interactions in the quantum system, the rate at which
these can be changed to assure convergence to the  desired final state
is determined by the adiabatic theorem~\cite{boh51}. Remarkably, we
will show that for the path corresponding to SA, the adiabatic
condition yields to the result obtained by Geman and Geman on the rate
of convergence to the optimal (ground) state of the classical
system~\cite{gem84}.

For simplicity, we study classical models defined on a lattice (or
graph), where a variable $\sigma_j=\pm1$ is defined on each site
(vertex) $j$, and is related with the states of a physical spin-1/2.
Any spin configuration of the $2^N$ possible ones is denoted as 
$[\sigma]\equiv [\sigma_1, \cdots, \sigma_N]$,  where $N$ is the total
number of sites (or problem size). An energy functional $E[\sigma]$
(cost function) is defined on the lattice and its value depends on the
state $[\sigma]$.  For example, in the Ising model, $E[\sigma]=
\sum_{ij} J_{ij} \sigma_i \sigma_j$, where two interacting spins $i$
and $j$ contribute $J_{ij}$ ($-J_{ij}$) to the energy if they are in
the same (different) state(s). In the canonical ensemble, the
expectation value of a thermodynamic variable $A$ at temperature $T$ is
given by
\begin{equation}
\label{correl1}
\langle A \rangle_T =\frac{1} {{\cal Z}(T)}
\sum \limits_{ [ \sigma ] } e^{ - \beta E{ [ \sigma ] }}
A_{ [ \sigma ] },
\end{equation}
where ${\cal Z}(T) = \sum \limits_{ [ \sigma ] } e^{ - \beta E{ [
\sigma ] }}$ is the partition function and $\beta=1/(k_B T)$, with
$k_B$ the Boltzmann's constant.

Any classical (finite-dimensional) spin model on a lattice can be
associated with a quantum one, defined on the same lattice, by mapping
every classical state $[\sigma]$ into a quantum state $\ket{[\sigma]}$.
In this way, the energy functional maps into a Hamiltonian operator
$H$. For spin-1/2 models, $H$ is given by mapping $\sigma_j \rightarrow
\sigma_z^j$ in $E[\sigma]$, where $\sigma_z^j$ is the Pauli operator
acting on the $j$th site. For example, $H=\sum_{ij} J_{ij} \sigma_z^i
\sigma_z^j$ in the Ising model. The $N$-spin (unnormalized) quantum
state $\ket{\psi(T)} = e^{- \beta H /2} \sum_{[\sigma]}
\ket{[\sigma]}$ (i.e., the Gibbs state), satisfies
\begin{equation}
\label{correl3}
\langle \hat{A} \rangle =\tr [\rho \hA]= \frac{\bra{\psi(T)} \hat{A}
\ket{\psi(T)}} {\bra{\psi (T)}\psi (T) \rangle} \equiv  \langle A
\rangle_T,
\end{equation}
where $\rho = e^{-\beta H}/{\cal Z}(T)$. The operator $\hat{A}$ is
determined by mapping the thermodynamic variable $A$, as described
above. Then, $[\hat{A},H]=0$.

The state $\ket{\psi(T)}$ can be shown to be the ground state of a
family of quantum Hamiltonians $H_q(T)$~\cite{hen04}, which are defined
on the same lattice. Each of these Hamiltonians can be connected
through a similarity transformation to a possible transition matrix
$M_q(T)$ of a Markovian process leading to the thermal distribution:
$H_q(T)= \one- e^{-\beta H/2} M_q(T) e^{\beta H/2}$. Interestingly, 
the interactions appearing in $H_q(T)$ are of comparable range to the
interactions of the classical model.  Therefore, a finite $T$ phase
transition of a $d-$dimensional classical system can then be identified
with a quantum phase transition of a $d-$dimensional quantum model.
Thus, constructing a specific $H_q(T)$ and studying its ground state
properties is of paramount importance as  it will allow us to build
different, yet more efficient algorithms to determine the thermodynamic
properties of the classical system.

We obtain the simplest form of $H_q(T)$  in the following way. First,
notice that the Pauli operator $\sigma_x^j $ (i.e., the spin-flip
operator acting on the $j$th site)  satisfies $\sigma_x^j e^{-\beta
H/2} \sigma_x^j = e^{\beta H_j} e^{-\beta H/2} \ , \  \forall j \in
[1,N]$. The Hamiltonian $H_j$ contains the terms in $H$ having the
operator $\sigma_z^j$ (i.e., the terms in $H$ that anticommute with
$\sigma_x^j$).  Moreover,   $\sigma_x^j \sum_{[\sigma]} \ket{[\sigma]}
=\sum_{[\sigma]} \ket{[\sigma]} $, and $H_q^j(T)\ket{\psi(T)}=0$, where
$H_q^j(T)=\sigma_x^j - e^{\beta H_j}$. In the basis  determined by the
states $\ket{[\sigma]}$, the off-diagonal elements of $H_q^j(T)$ are
non-negative, and the coefficients appearing in $\ket{\psi(T)}$ are all
positive.  The Perron-Frobenius theorem~\cite{hor85} guarantees then
that for $T >0$, $\ket{\psi(T)}$  is the unique ground state of the
irreducible quantum  Hamiltonian  $H_q(T) = -\chi \sum_j H_q^j(T)$. The
coefficient $\chi = e^{-\beta p}$, with $p \approx \max_j |H_j| = {\cal
O}(1)$, is set for normalization purposes in order to satisfy
$|H_q(T\rightarrow 0) | < \infty$. At this point, we would like to
emphasize the simplicity of our particular mapping: The thermodynamic
properties of any spin-1/2 classical system can be  obtained by
studying the ground state properties of  a spin-1/2 quantum model, with
{\em classical} interactions determined by $T$ and $H$ (i.e., the
classical system), and an external (homogeneous) transverse field of
magnitude $\chi$. Remarkably, this field generates quantum fluctuations
that are in one-to-one correspondence with the classical fluctuations
at temperature $T$.  In particular,   $H_q(T \rightarrow \infty)
\approx (N -\sum_j \sigma_x^j)$, so its ground state has all spins
aligned along the external field, i.e. $\ket{\psi(T \rightarrow
\infty)} \approx \sum_{[\sigma]} \ket{[\sigma]}$. This quantum state 
can be identified with the completely mixed state in the classical
model.  In the limit of low $T$  we obtain $H_q(T \sim 0) \approx \chi
\sum_j e^{\beta H_j}$, whose expectation value is minimized by the
ground state(s)  of the classical model, i.e. $\ket{\psi(T \sim 0)}$ is
also a lowest energy state of $H$.

To illustrate these results, we consider the homogeneous
one-dimensional Ising model $H= J \sum_{i=j}^N \sigma_z^j
\sigma_z^{j+1}$. In this case,  $H_q^j (T)= \sigma_x^j - x^2 -xy
(\sigma_z^{j-1} \sigma_z^j + \sigma_z^j \sigma_z^{j+1}) - y^2
\sigma_z^{j-1} \sigma_z^{j+1}$, with $x=\cosh (\beta J)$, and $y=\sinh
(\beta J)$. The Hamiltonian $H_q(T)$ denotes then a frustrated quantum
Ising model, with next-nearest-neighbor interactions, and a transverse
magnetic field of magnitude $\chi$. Such a frustration forbids the
existence of an ordered quantum phase unless $T \rightarrow 0$. This
result is related to the non-existence of an ordered phase at finite
temperature in the classical model.

Within our context, we can interpret the SA procedure as a (real time)
quantum evolution where we start from the initial  quantum state
$\ket{\psi(T \rightarrow \infty)}\approx \sum_{[\sigma]}
\ket{[\sigma]}$, and next we decrease the interaction parameter $T(t)$
(related to the temperature of the classical model) in $H_q(T)$. If
such an evolution is performed adiabatically, we remain in the desired
ground state $\ket{\psi(T(t))}$ at any time $t$. Therefore, the gap
$\Delta(T)$ between the ground and first excited states of $H_q(T)$
plays an important role on the rate at which $T(t)$ must be decreased.
This gap can be shown to satisfy $\Delta(T) \ge 2 \sqrt{2 \pi N}
e^{-(\beta p +1)N} = \bar{\Delta}(T)$. Such a lower bound can be
determined using the inequalities in Ref.~\cite{hop63} and considering
that $(N-H_q(T))^N$ is a strictly positive operator. It is based on the
worst-case scenario (i.e., for the most general form of $H$), so it is
expected to be improved depending on the nature of the interactions of
the classical system, such as translational invariance. The rate of the
evolution is then determined by the adiabatic condition~\cite{boh51}
\begin{equation}
\label{adiabcon}
  \max_{m} \left| \frac{ \bra{\psi_m(T(t))} \partial_T H_q(T)
\ket{\psi(T(t))} } {\Delta^2_m(T(t)) \sqrt{{\cal Z}(T(t))} }\ 
\partial_t T \right|= \epsilon,  \ 0  \le t \le {\cal T},
\end{equation}
where $\epsilon$ determines an upper bound to the probability of
finding the system in any other (normalized) excited eigenstate
$\ket{\psi_m(T)}$ of $H_q(T)$,  $\Delta_m(T)$ is the energy gap between
$\ket{\psi_m(T)}$ and $\ket{\psi(T)}$ (e.g., $\Delta_1(T) \equiv
\Delta(T)$), and $\cal T$ is the total time of the evolution. The {\em
lhs} of Eq.~(\ref{adiabcon}) can be bounded above by $pN[2 k_B T^2
\bar{\Delta}(T)]^{-1} |\partial_ t T|$. To see this, note that
\begin{equation}
 \partial_T H_q(T) \ket{\psi(T)} \equiv [\partial_T (-\beta H/2),
H_q(T) ]  \ket{\psi(T)} ,
\end{equation}
as $-\beta H/2$ generates the translations of  $\ket{\psi(T)} $.
Therefore,
\begin{equation}
\label{sa-scal}
\frac{|\bra{\psi_m(T)}\partial_T H_q(T) \ket{\psi(T)}|} {\Delta_m(T)} 
=\frac{ |\bra{\psi_m(T)}H \ket{\psi(T)}| }{(2k_B T^2)},
\end{equation}
with $|\bra{\psi_m(T)}H \ket{\psi(T)}| \le pN\sqrt{{\cal Z}(T)}$. This
upper bound is not necessarily tight. Equation~(\ref{sa-scal}) implies
a resource requirement of ${\cal T} \approx {\cal O} [1/\epsilon
\bar{\Delta}(T)]$ instead of ${\cal T} \approx {\cal O} [1/\epsilon
\bar{\Delta}^2(T)]$, which is the common resource scaling associated
with an adiabatic evolution. [Nevertheless, both scalings will yield to
similar asymptotic behavior for $T(t)$.] Integrating
Eq.~(\ref{adiabcon}), replacing $\min_m [\Delta_m(T(t))]$ by
$\bar{\Delta}(T(t))$, yields to 
\begin{equation}
\label{tbehav}
T(t) \approx \frac{pN}{k_B\log( \alpha t+1) }, \ 0 < t \le {\cal T},
\end{equation}
where $\alpha$ decreases exponentially with the system size $N$ and is
proportional to $\epsilon$,  and $T({\cal T})$ is the temperature at
which we want to study classical system.  That is, if $T$ is decreased
as given by Eq.~(\ref{tbehav}), convergence to the desired state is
guaranteed. In the limit $T({\cal T}) \rightarrow 0$ and $\log t \gg N
\gg 1$, we obtain $T(t) \approx (pN)/(k_B \log t)$ which agrees with
the asymptotic convergence rate obtained in Ref.~\cite{gem84} for SA.
Such an agreement relies on the fact that the energy gap of $H_q(T)$ is
also the energy gap of the transition matrix $M_q(T)$, which is known
to determine the {\em mixing} time (or time required to reach thermal
equilibrium) ${\cal T}_M \approx {\cal O}(1/\Delta(T))$. That is, in
the SA scheme one never departs from equilibrium if the temperature is
decreased with the above convergence rate. Equivalently, in our
context, the overlap between the adiabatically evolved quantum state
and $\ket{\psi(T(t))}$ is always close to 1. Note that Eq.~(\ref{tbehav})
holds even if the interactions in $H$ are of long-range nature.

QA has been proposed in Ref.~\cite{kad98} as an alternative method to
reach the optimal (ground) state of a classical system with Ising-like
interactions. Contrary to SA,  the time-dependent quantum state in QA
does not correspond, in general, to a thermal configuration of the
original classical model.   In this case, the quantum model Hamiltonian
is given by $H'_q(\gamma) = H -\gamma \sum_j \sigma_x^j$, where
$\gamma$ is decreased from a very large value, corresponding to $T
\rightarrow  \infty$, to $\gamma\approx 0$, corresponding to $T \approx
0$.  If $\gamma$ is slowly (adiabatically) changed, this method also
allows us to reach the ground state of $H$. Similar techniques have
been proposed to study the complexity of solving NP-complete problems,
such as 3-SAT, using a quantum computer (QC)~\cite{far01}. Numerical
and analytical results show that, for certain optimization problems, QA
might enable a faster convergence rate to the optimal state than
SA~\cite{kad98,suz05,mor06}. Faster convergence of QA could be
attributed to  a decrease in the probability of driving the classical
system to a local minima, as its dimension  is effectively increased by
one.  Nevertheless, it has also been observed that in some
cases~\cite{far02} QA performs similarly to SA.  Note, however, that
one could construct different Hamiltonian paths to approach the optimal
state. Each path yields to a particular convergence rate that has to be
determined on a case by case basis.

Using the classical-quantum mapping described above, the QA method can
be extended to simulate classical statistical mechanics. To show this,
we define a quantum Hamiltonian $\tilde{H}_q(\gamma) = \chi \sum_j
e^{\beta H_j} - \gamma \sum_j \sigma_x^j$, having $\ket{\psi(\gamma)}$ and
$\ket{\psi_m(\gamma)}$ as ground and excited states.
Here,  $\gamma$ is
adiabatically decreased from a very large value towards $\gamma \approx
\chi$. In this way,  the initial state $\sum_{[\sigma]} \ket{[\sigma]}$
is transformed into the desired state $\ket{\psi(T)}$. Notice that,
from our viewpoint, QA differs from SA only by the choice of path used
to reach the desired state. To successfully implement this annealing
procedure,  the rate at which $\gamma$ must be decreased is determined
by the adiabatic condition,  i.e. by the gap $\Delta(\gamma)$ between
the ground and first excited states of ${\tilde H}_q(\gamma)$. This gap can
be shown to satisfy $\Delta(\gamma) \ge 2 \sqrt{2 \pi N} e^{-N}
(1+c)^{-N} \gamma^N = \bar{\Delta}(\gamma)$~\cite{hop63}, with $\gamma < c$. 
Like the SA
case, and for the worst-case scenario, 
the adiabatic condition~\cite{boh51} yields to 
\begin{equation}
\label{bbehav}
\gamma(t) \approx [(2N-1)(\bar{\alpha} t )]^{-1/(2N-1)},  \ 0 <
t \le {\cal T},
\end{equation}
where $\bar{\alpha}$ depends on $N$, $c$, and $\epsilon$,  and $\gamma({\cal
T})=\chi$ is determined by $T$. 
In the limit $\log
t \gg N \gg 1 $, and $\gamma({\cal T}) \ll 1$, we obtain $\gamma(t)
\approx (2N \bar{\alpha} t)^{(-1/2N)}$. If $|\bra{\psi_m(\gamma)}
\partial_\gamma \tilde{H}_q(\gamma) \ket{\psi (\gamma)}| \le x
\Delta_m(\gamma)$, with $\Delta_m(\gamma)$ the corresponding energy gap  and $x\approx{\cal
O}(N^q)$, the coefficient $2N$ in Eq.~(\ref{bbehav}) can then be
replaced by $N$. In this manner, the convergence rate  is in agreement
with the result obtained in Ref.~\cite{mor06}. Note, however, that this
annealing schedule does not provide an advantage with respect to SA as
$\gamma$ must be decreased to $\gamma({\cal T})= \chi$, which is
exponentially small in $1/T$.

The QA procedure to simulate $T >0$ can be directly implemented on a CC
\cite{Note1}. If the path-integral Monte Carlo method is chosen to
simulate a $d=1$ Ising-like model with nearest-neighbor interactions,
$\tilde{H}_q(\gamma)$ has to be mapped onto the $2$-dimensional
classical model, with energy functional
\begin{equation}
\label{montecarlo}
\bar{E}[\sigma]= \frac{\tilde{\beta}}{L} \sum_{k=1}^L \sum_{ij}
\tilde{J}_{ij}(\beta) \sigma_{ik} \sigma_{jk} + \xi(\beta,t) \sum_{k=1}^L
\sum_{i=0}^N \sigma_{ik} \sigma_{i(k+1)}.
\end{equation}
Here, $[\sigma]=[\sigma_{11},\sigma_{21},\cdots,\sigma_{NL}]$ is one of
the $2^{N+L}$ possible spin configurations, and $\sigma_{ik}=\pm 1$.
The parameter $L$ denotes the number of copies of the system in the
extra dimension (i.e., the Trotter discretization) and satisfies $L \gg
1$. The coupling constants $\tilde{J}_{ij}(\beta)$ are defined via
$\chi \sum_j e^{\beta H_j} \equiv \Lambda(\beta) + \sum_{ij}
\tilde{J}_{ij}(\beta) \sigma_z^i \sigma_z^j$,  with
$\tilde{J}_{ij}(\beta \rightarrow 0) \approx 0$. The coefficient
$\tilde{\beta}$ is given by the effective temperature of the quantum
system and is not related with the temperature at which the classical
system is studied. Therefore, $\tilde{\beta} \gg 1$ and
$\tilde{\beta}/L = \delta \tau$, with $\delta \tau$ being the
time-slice of the discretization. The (ferromagnetic) coupling between
two adjacent copies is determined by  $\xi(t)= \log[\coth
(\tilde{\beta} \chi \gamma(t)/L)]/2$, and its magnitude increases as
the transverse field $\gamma(t)$ decreases to $\gamma({\cal T})= \chi$,
determined by $T$.  In order to simulate more general classical systems
at finite temperatures, the interactions appearing in
Eq.~(\ref{montecarlo}) must be modified accordingly.

Note that the classical-quantum mapping can be extended and used to
study any (finite-dimensional) classical system other than Ising-like
models. In particular, it can be extended to simulate $s$-spin
classical systems ($s > 1/2$), where a variable $\sigma_j=[-s,-s+1,
\cdots,s-1, s]$ is defined on each site $j$. In the case of QA, the
ground state of $\tilde{H}_q^s(\gamma)=\chi \sum_j e^{\beta H_j} -
\gamma \sum_j X_j $ will determine the statistical properties of the
classical model when $\gamma \rightarrow \chi$. The operators $X_j \in
\fsu(2s+1)_j$  satisfy $[X_j,S_z^i ]=0 ,\ \forall i \ne j,$ and $X_j
S_z^j= - S_z^j X_j$, with $X_j^2=\one$. Here,  $S_z^j \in \fsu(2)_j$ is
the angular momentum operator along the $z$-axis and determines $H_j$.
In matrix representation, $X$ has 1's in the anti-diagonal  and 0's
otherwise. For example, in the $s=1$ (three-state) Potts
model~\cite{pot52}, $E[\sigma] = J\sum_j \delta(\sigma_j,\sigma_{j+1})$
and $H=J/2 \{ \sum_j  [S_z^j S_z^{j+1}(1+S_z^j S_z^{j+1})]-
2[(S_z^j)^2+ (S_z^{j+1})^2] \} $. Therefore,  $H_j = J/2 [ S_z^{j-1}
S_z^j + S_z^j S_z^{j+1}]$ and $X_j = 1-(S_z^j)^2+[(S_+^j)^2+
(S_-^j)^2]/2 $, defining the corresponding $\tilde{H}^{s=1}_q(\gamma)$.
The annealing schedule is again determined by adiabatically changing
$\gamma(t)$ from a large value, where the initial state of the system is
$\sum_{[\sigma]} \ket{[\sigma]}$, to $\gamma({\cal T}) = \chi=e^{-\beta
p}$,  where the final state of the system is $e^{-\beta H/2} \sum_{[\sigma]}
\ket{[\sigma]}$.

It is important to stress that one can easily implement this {\em
extended} QA (EQA) procedure by using  current numerical methods. 
Since our analysis has only focused on the worst-case scenario,  we
would expect that for certain problems EQA should outperform
SA~\cite{kad98}. Moreover, one can always design other   annealing
procedures than the ones we have described. This can be done  by
constructing other quantum Hamiltonians having $\ket{\psi(T)}$ as their
ground state, and by introducing an extra interaction that is  slowly
changed to converge to the desired state.  Depending on the path
considered, it is expected a different behavior for the way that the
relevant energy gap closes, and a different convergence rate as
determined by the adiabatic condition.

So far, we have considered that the  lower bound  in the gap is
exponentially small in $\beta N$, for SA, or $N \log \gamma$, for QA. 
One may wonder what the convergence rate for $T(t)$ or $\gamma(t)$ is,
when the gap can be bounded below by $(\beta N)^{-1/q}$ or $(N \log
\gamma)^ {-1/q}$, with $q \ge 0$ independent of $N$ and $\beta$. In
this case, integration of Eq.~(\ref{adiabcon}) yields to a convergence
rate for SA of $T(t) \approx {\cal O} [\alpha / t^{1/(q+1)}]$, with
$\alpha$ a constant that depends on $N$ and $\epsilon$. This is a much
faster convergence rate than the one obtained in Eq.~(\ref{tbehav}).

In this Letter, we have shown how to simulate the thermodynamic
properties of an arbitrary classical model in $d$ dimensions by
studying the ground state properties of a $d$ dimensional quantum
system. This was achieved by an exact classical-quantum mapping. We
have used the adiabatic theorem of quantum mechanics to analyze the
convergence rate and resources required to reach the corresponding
ground state. Our approach provides a unifying framework to address on
an equal footing the well-known optimization methods of simulated and
quantum annealing. These annealing procedures can be understood as two
different evolution paths of the quantum system. It is remarkable, that
the  annealing rates obtained by using the adiabatic condition are in
agreement with previous known results~\cite{gem84,mor06}, which were
obtained in the context of stochastic approaches such as path-integral
or Green's function Monte Carlo. It is expected, however, that a QC 
will require less resources (e.g., quadratic speed-up) than a CC to
solve these optimization problems.  This issue will be addressed
elsewhere.

%What happens in a critical point with conformal invariance?}

\acknowledgments
We are thankful to E. Knill, F. Verstraete,  H. Barnum, and H.
Nishimori for useful discussions. This work was carried out under the
auspices of the National Security Administration of the US DOE at LANL
under Contract No. DE-AC52-06NA25396.

\end{document}